\documentclass[a4paper,11pt]{article}
\usepackage{pos}
\usepackage{braket}

\newcommand*{\cf}{cf.\ }
\newcommand*{\eg}{e.\,g.\ }
\newcommand*{\ie}{i.\,e.\ }

\title{NNLO QCD corrections to $B$-meson mixing}

\author*[a]{Vladyslav Shtabovenko}

\affiliation[a]{Institut f{\"u}r Theoretische Teilchenphysik,
Karlsruhe Institute of Technology (KIT),\\
76128 Karlsruhe, Germany}

\emailAdd{v.shtabovenko@kit.edu}

\abstract{We report on the calculation of next-to-next-to-leading order (NNLO) QCD corrections to the width difference $\Delta \Gamma_s$ in the neutral $B$-meson mixing process $B^0_s - \bar{B}^0_s$. These contributions represent an important step in the task of reducing the existing large perturbative errors in the theory prediction for $\Delta \Gamma_s$ and approaching the current experimental uncertainties. We explain the theoretical framework employed in this computation and point out important subtleties 
in the treatment of evanescent operators and the renormalization. Part of our new results is already available in the literature, while the remaining pieces are expected to be published later this year.
}

\FullConference{%
  *** The European Physical Society Conference on High Energy Physics (EPS-HEP2021), ***\\
  *** 26-30 July 2021 ***\\
  *** Online conference, jointly organized by Universität Hamburg and the research center DESY ***
}

\begin{document}
\maketitle

\section{Introduction}	

In view of current null searches for new physics at the LHC, 
the increasing number of anomalies in the flavor sector of the Standard Model
(\cf \eg \cite{Lenz:2021bkv} for a recent overview) hints that the much anticipated
beyond the Standard Model (BSM) particles might indeed be discovered 
at the precision and not at the high-energy frontier. Amid the well justified efforts of the BSM community to explain such 
anomalies by developing suitable models and studying their implications, 
one should not forget about the existence of flavor precision observables
and their importance for our present understanding of the SM. The word
``precision'' indicates that such observables are equally well accessible to
experimental measurements and theoretical calculations, which creates a friendly
competition between these two particle physics communities.

One of such observables is the width difference $\Delta \Gamma_s$ that arises
in the $B^0_s - \bar{B}^0_s$ oscillations. This time-dependent process can
be described via
\begin{equation}
	i \frac{d}{d t}  \begin{pmatrix}  \ket{B^0_s  (t)} \\ \ket{\bar{B}^0_s (t)} \end{pmatrix} = \left ( \hat{M} - \frac{i}{2} \hat{\Gamma} \right ) \begin{pmatrix} \ket{B^0_s  (t)} \\ \ket{\bar{B}^0_s (t)} \end{pmatrix} \label{eq:mixing},
\end{equation}
where $\hat{M}$ and $\hat{\Gamma}$ are Hermitian matrices. In the absence of mixing $\hat{M}$ and $\hat{\Gamma}$ would have no off-diagonal elements, while their diagonal entries would correspond to the $B^0_s$ meson mass and width respectively.
Yet by exchanging $W$-bosons in box diagrams, $b$- and $\bar{s}$-quarks can turn into $\bar{b}$ and $s$, which corresponds
to the flavor eigenstate $B_s^0$ transforming into its antiparticle and back via weak interactions.
This loop-suppressed flavor changing neutral current induces nonzero values of $M_{12}$ and $\Gamma_{12}$.
Solving Eq.\,\eqref{eq:mixing} one arrives at two mass eigenstates $\ket{B_{L}}$ (lighter, almost CP-even) and $\ket{B_{H}}$ (heavier, almost CP-odd)
\begin{equation}
	\ket{B_{L/H}} = p \ket{B^0_s} \pm q \ket{\bar{B}^0_s}, \qquad \text{with } p^2+q^2 = 1.
\end{equation}
The mass and lifetime differences between these eigenstates
are given by
\begin{equation}
	\Delta m_s	 \equiv  M_H - M_L  = 2 |M_{12}|, \quad \Delta \Gamma_s	 \equiv \Gamma_L - \Gamma_H    \approx 2 |\Gamma_{12}|.
\end{equation}
The width difference $\Delta \Gamma_s$ is of limited sensitivity to new physics and would deviate from its SM value only 
if BSM particles contributing through loops are light and weakly coupled to SM fields. In fact, $\Delta m_s$ and $\Delta \Gamma_s$ are complementary probes of new physics because of the former quantity probing effects of heavy (multi-TeV-mass) particles.

This implies that $\Delta \Gamma_s$ is a superb probe for our understanding of the SM and higher perturbative corrections thereto. Indeed, in the past years experimentalists have done a great job on reducing the statistical and systematic errors
in their measurements \cite{LHCb:2019nin,CMS:2020efq,ATLAS:2020lbz} of  $\Delta \Gamma_s$ and achieving per cent level precision 
\cite{HFLAV:2019otj} with  
\begin{equation}
	\Delta \Gamma^{\textrm{exp}}_s = (0.085 \pm 0.004) \textrm{ ps}^{-1}
\end{equation}
as compared to the current theory predictions \cite{Beneke:1998sy,Ciuchini:2001vx,Ciuchini:2003ww,Lenz:2006hd,Asatrian:2017qaz,Asatrian:2020zxa} 
\begin{align}
	\Delta \Gamma_s^{\overline{\text{MS}}} &= (0.088 \pm 0.011{}_{\textrm{pert.}} \pm 0.002{}_{B,\tilde{B}_S} \pm 0.014_{\Lambda_{\textrm{QCD}}/m_b} ) \textrm{ ps}^{-1}, \label{eq:exp1} \\
	\Delta \Gamma_s^{\text{pole}} &= (0.077 \pm 0.015{}_{\textrm{pert.}} \pm 0.002{}_{B,\tilde{B}_S} \pm 0.017_{\Lambda_{\textrm{QCD}}/m_b} ) \textrm{ ps}^{-1},  \label{eq:exp2}
\end{align}
that suffer from large perturbative uncertainties (denoted as ``pert.''). The two values given in Eqs.\eqref{eq:exp1} and \eqref{eq:exp2} correspond to different renormalization schemes. The reduction of these uncertainties necessitates an inclusion of the missing QCD corrections to the $B^0_s - \bar{B}^0_s$ mixing at 2- and 3-loop level, which is the main task of our project.

\section{Calculation}

In the calculation of $\Delta \Gamma_s$ we work with the $|\Delta B|=1$ effective Hamiltonian
that can be expressed using the following set of operators \cite{Chetyrkin:1997gb}
\begin{align}
	\mathcal{H}_{\textrm{eff}}^{|\Delta B|=1} 
	&=   \frac{4G_F}{\sqrt{2}}  \left[
	-\, \lambda^s_t \Big( \sum_{i=1}^6 C_i Q_i + C_8 Q_8 \Big) 
	- \lambda^s_u \sum_{i=1}^2 C_i (Q_i - Q_i^u) \right. \\ \nonumber
	& \phantom{\frac{4G_F}{\sqrt{2}} \Big[}
	\left.
	+\, V_{us}^\ast V_{cb} \, \sum_{i=1}^2 C_i Q_i^{cu} 
	+ V_{cs}^\ast V_{ub} \, \sum_{i=1}^2 C_i Q_i^{uc} 
	\right]
	+ \mbox{h.c.},
\end{align}
where $V_{ij}$ stand for the CKM matrix elements and $\lambda^s_a = V_{as}^\ast V_{ab}$
denote products thereof, while $C_i$ are matching coefficients arising the matching between SM and $\mathcal{H}_{\textrm{eff}}^{|\Delta B|=1}$. The basis also includes so-called evanescent operators $E[Q_i]$ \cite{Dugan:1990df,Herrlich:1994kh} 
that are formally of order $\mathcal{O}(\varepsilon)$ and therefore vanish in the limit $d \to 4$.
Their relevance arises from the fact that certain relations from the 4-dimensional 
Dirac algebra such as Fierz identities or the Chisholm identity cannot be translated to $d$ 
dimensions in a unique fashion.  The proper handling
of the evanescent operators during the renormalization and in the matching is one of the
conceptual challenges accompanying this calculation. Explicit definitions of all $|\Delta B|=1$ operators
entering our matching calculations at 2 and 3 loops can be found in \cite{Gerlach:2021xtb} and \cite{Gerlach:2021prep3L}
respectively.

In quantum field theory $\Delta\Gamma_s \approx 2 |\Gamma_{12}|$ is related to $\Gamma_{12}$,
the absorptive part of a bilocal matrix element featuring a time-ordered product of two 
$|\Delta B|=1$ effective Hamiltonians. Simplifying this quantity by means of the Heavy
Quark Expansion (HQE) \cite{Khoze:1983yp,Shifman:1984wx,Khoze:1986fa,Chay:1990da,Bigi:1991ir,Bigi:1992su,Bigi:1993fe,Blok:1993va,Manohar:1993qn} we arrive at \cite{Lenz:2006hd}
\begin{equation}
	\Gamma_{12} = - (\lambda_c^s)^2\Gamma^{cc}_{12} 
	- 2\lambda_c^s\lambda_u^s \Gamma_{12}^{uc} 
	- (\lambda_u^s)^2\Gamma^{uu}_{12} 
	\,,
\end{equation}
with
\begin{equation}
	\Gamma_{12}^{ab} 
	= \frac{G_F^2m_b^2}{24\pi M_{B_s}} \left[ 
	H^{ab}(z)   \langle B_s|Q|\bar{B}_s \rangle
	+ \widetilde{H}^{ab}_S(z)  \langle B_s|\widetilde{Q}_S|\bar{B}_s \rangle
	\right] + \mathcal{O} (\Lambda_{\text{QCD}}/m_b), \label{eq:DB2}
\end{equation}
where $z \equiv m_c^2/m_b^2$. The determination of the relevant QCD corrections to
the Wilson coefficients $H^{ab}(z)$ and $\widetilde{H}^{ab}_S(z)$ is the main goal of our project.
The $|\Delta B|=2$ operators appearing in Eq.\,\eqref{eq:DB2} are defined as
\begin{equation}
	Q = \bar{s}_i \gamma^\mu \,(1-\gamma^5)\, b_i \; \bar{s}_j \gamma_\mu
	\,(1-\gamma^5)\, b_j\,, \qquad  
	\widetilde{Q}_S = \bar{s}_i \,(1-\gamma^5)\, b_j\; \bar{s}_j \,(1-\gamma^5)\,
	b_i, \label{eq:opDB2}
\end{equation}
with $i,j$ specifying the color indices of the quark fields. The complete $|\Delta B|=2$ operator basis (\cf \cite{Gerlach:2021xtb})
also features suitable evanescent operators as well as the operator
$R_0$ whose renormalized matrix elements are $1/m_b$-suppressed, while its bare matrix elements are not \cite{Beneke:1998sy}.

\begin{figure}[t]
	\begin{center}
		\begin{tabular}{cccc}
			\includegraphics[width=0.25\textwidth]{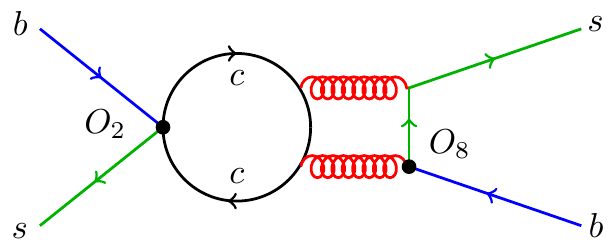} &
			\includegraphics[width=0.21\textwidth]{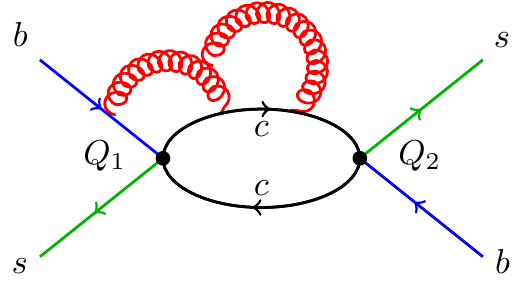} &
			\includegraphics[width=0.15\textwidth]{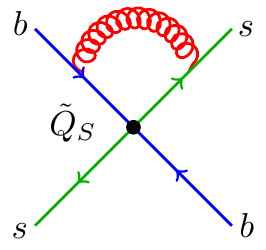} &
			\includegraphics[width=0.15\textwidth]{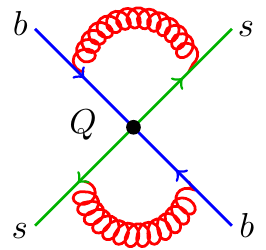} \\
			(a) & (b) & (c) & (d)
		\end{tabular}
	\end{center}
	\caption{\label{fig:sample}Sample $\Delta B=1$ and $\Delta B=2$ Feynman diagrams contributing to the
		process $b  \bar{s} \to \bar{b}  s$. Here (a) and (b) represent $\Delta B=1$ contributions
		from $Q_{2} \times Q_{8}$ (2 loops) and $Q_{1} \times Q_{2}$ (3 loops), while (c)
		and (d) visualize 1- and 2-loop matrix elements of $\Delta B=2$ operators $\tilde{Q}_s$ and $Q$.}
\end{figure}

In the matching between $|\Delta B|=1$ and $|\Delta B|=2$ effective theories we choose to treat the $s$ quark as massless 
and to set its external momentum to zero, while the external momentum of the $b$ quark is taken on-shell. Furthermore, at 2 loops
we expand in $z$ up to $\mathcal{O}(z)$, while the 3-loop diagrams are evaluated
in the $z =0$ limit. On the $|\Delta B|=1$ side of the matching we calculate all possible operator insertions up to 
2 loops \ie all combinations of $Q_{1,2}$, $Q_{3-6}$ and $Q_{8}$ appearing in either of the two vertices. Notice that the
2-loop contribution to $Q_{8} \times Q_{8}$ actually belongs to NNNLO, but here we obtain it
as a byproduct of our calculation. At 3 loops
we evaluate only the $Q_{1,2} \times Q_{1,2}$ correlator. As far as the 
$|\Delta B|=2$ theory is concerned, the 2-loop $|\Delta B|=1$ contributions are matched to
the 1-loop $|\Delta B|=2$ diagrams, while the 3-loop $Q_{1,2} \times Q_{1,2}$ correlator requires
us to consider 2-loop corrections to the $|\Delta B|=2$ operators.  Some of the representative Feynman 
diagrams visualizing the corresponding operator insertions are shown in Fig.\,\ref{fig:sample}.

\section{Technical details}

To carry out the analytic evaluation of the Feynman diagrams on 
both sides of the matching we make use of our well tested 
in-house calculational setup. We generate the required Feynman graphs using
\textsc{Qgraf} \cite{Nogueira:1991ex} and employ \textsc{q2e/exp} \cite{Harlander:1998cmq,Seidensticker:1999bb} or 
\textsc{tapir} \cite{Gerlach:tapir} to insert Feynman rules and identify
the occurring integral topologies. The resulting amplitudes are then
processed with the aid of the \textsc{FORM}-based \cite{Kuipers:2012rf} 
\textsc{calc} framework. For cross checks of the results
obtained from single diagrams we also employ \textsc{FeynRules} \cite{Alloul:2013bka},
\textsc{FeynArts} \cite{Hahn:2000kx} and \textsc{FeynCalc} \cite{Mertig:1990an,Shtabovenko:2016sxi,Shtabovenko:2020gxv}. The latter, in conjunction
with \textsc{Fermat} \cite{Lewis:Fermat} is also used to derive tensor integral
reduction formulas \cite{Pak:2011xt} that are used in our \textsc{FORM} code. Alternatively, we also employ a set of suitable Dirac and color projectors. \textsc{FIRE} \cite{Smirnov:2019qkx} and \textsc{LiteRed} \cite{Lee:2013mka} allow us to IBP-reduce \cite{Chetyrkin:1981qh,Tkachov:1981wb} the occurring loop 
integrals to a small set of master integrals, many of which have already
been calculated in the past \cite{Smirnov:2012gma,Fleischer:1999hp}. Most of the on-shell 3-loop
master integrals, however, appear to be new and need to be calculated 
from scratch. This is done using \textsc{FeynCalc}, \textsc{HyperInt} \cite{Panzer:2014caa},
\textsc{HyperLogProcedures} \cite{Schnetz:HLP} and \textsc{PolyLogTools} \cite{Duhr:2019tlz}, so that at the end of the day we are able to obtain explicit analytic results for all integrals occurring in this matching calculation. We also  
cross check these results numerically using \textsc{pySecDec} \cite{Borowka:2017idc,Borowka:2018goh,Heinrich:2021dbf} and
\textsc{FIESTA} \cite{Smirnov:2015mct}.

\section{Renormalization and matching}

We renormalize the bare $|\Delta B|=1$ and $|\Delta B|=2$ amplitudes 
in the $\overline{\textrm{MS}}$ scheme. Notice that in addition to the QCD 
renormalization constants we also need to take into account the renormalization
of $|\Delta B|=1$ and $|\Delta B|=2$ operators. The generic operator renormalization 
matrix is of the form
\begin{eqnarray}
	Z = \begin{pmatrix} Z_{QQ} & Z_{QE} \\ Z_{EQ} & Z_{EE} \end{pmatrix},
\end{eqnarray}
where the submatrices $Z_{ij}$ indicate mixing between different operator
subclasses. For example, $Z_{EQ}$ encodes the mixing of evanescent
into physical operators at order $\mathcal{O}(\alpha_s)$.
The $|\Delta B|=1$ renormalization matrix is readily available in the literature
\cite{Gambino:2003zm} and is sufficient for NNLO accuracy. As far as
$Z^{|\Delta B|=2}$ is concerned, the situation is less favorable, where mostly only
$Z_{QQ}$ can be found in the literature. For this reason we choose to determine 
$Z^{|\Delta B|=2}$ tailored to our operator basis at 2 loops in a separate calculation.

To ensure the correctness of our matching calculation, 
at 2 loops we not only regularize UV and IR divergences dimensionally,
but also employ a finite gluon mass as an infrared regulator.
The latter makes the evaluation of the amplitudes and the calculation of the master
integrals somewhat more involved but leads to significant simplifications in the matching. In particular, upon the UV renormalization our $|\Delta B|=1$ and $|\Delta B|=2$ amplitudes are free of $\varepsilon$ poles, so that one can safely take the limit $d \to 4$, where the matrix elements of evanescent operators vanish.

If we choose to work with massless gluons and therefore use the same $\varepsilon$ as our UV and IR regulator, then the renormalized amplitudes on both sides of the matching still contain IR poles and the contributions of evanescent operators must be kept. In this case the matching should be carried out according to the prescriptions outlined in \cite{Ciuchini:2001vx}. 
This way all IR poles cancel and at 2-loops we obtain the same matching coefficients as in the calculation
with massive gluons. At 3 loops we only work with massless gluons and using the method from \cite{Ciuchini:2001vx} we observe an explicit cancellation of all IR poles in the matching.

\section{Results}

Due to the large number of new matching coefficients obtained in the course of this project,
we summarize the obtained results in Table \ref{tab:results}, which also indicates the previous status quo
from the literature. Since our 2- and 3-loop results are of $\mathcal{O}(z)$
and $\mathcal{O}(z^0)$ respectively, it is understood that when comparing to the literature
we also need to expand the relevant expressions in $z$. Under these conditions we confirm all
the existing literature results, including the fermionic part of the 3-loop correlator 
$Q_{1,2} \times Q_{1,2}$ computed in \cite{Asatrian:2017qaz}.
\begin{table}[!h]
	\begin{center}
		\begin{tabular}{|c|c|c|}
			\hline
			Contribution & Literature result & This work \\
			\hline
			$Q_{1,2} \times Q_{3-6}$ & 2 loops, $z$-exact, $n_f$-part only \cite{Asatrian:2020zxa} & 2 loops, $\mathcal{O}(z)$, full \\	\hline
			$Q_{1,2} \times Q_{8}$  & 2 loops, $z$-exact, $n_f$-part only \cite{Asatrian:2020zxa} &  2 loops, $\mathcal{O}(z)$, full \\	\hline
			$Q_{3-6} \times Q_{3-6}$ & 1 loop, $z$-exact, full \cite{Beneke:1996gn}   &  2 loops, $\mathcal{O}(z)$, full\\	\hline
			$Q_{3-6} \times Q_{8}$  & 1 loop, $z$-exact, $n_f$-part only \cite{Asatrian:2020zxa}   &  2 loops, $\mathcal{O}(z)$, full\\	\hline
			$Q_{8} \times Q_{8}$ & 1 loop, $z$-exact, $n_f$-part only \cite{Asatrian:2020zxa}   & 2 loops, $\mathcal{O}(z)$, full\\
			$Q_{1,2} \times Q_{1,2}$ & 3 loops, $\mathcal{O}(\sqrt{z})$, $n_f$-part only \cite{Asatrian:2017qaz}   & 3 loops, $\mathcal{O}(z^0)$, full \\
			\hline
		\end{tabular}
		\caption{Overview of the existing and new results required for the NNLO theory prediction of  $\Delta\Gamma_s$ that were considered in this work. With ``$n_f$-part only'' we signify that the corresponding literature result provides only fermionic contributions, while ``full'' means that both fermionic and nonfermionic pieces are included.}
		\label{tab:results}
	\end{center}
\end{table}
The final NNLO theory prediction for the width difference is still work in progress, 
due to additional checks required to ensure that our treatment of the $1/m_b$-suppressed $R_0$ operator at 3 loops
is correct. As far as the 2-loop contributions are concerned, explicit results for the 
$Q_{1,2} \times Q_{3-6}$ diagrams have already been published in \cite{Gerlach:2021xtb}, while 
the remaining 2-loop matching coefficients are expected to appear soon \cite{Gerlach:2021prep2L}. The 3-loop result together
with the updated NNLO theory prediction are also in preparation \cite{Gerlach:2021prep3L}.

To highlight the relevance of our computation, let us observe that alone the complete (\ie not just its fermionic piece) 2-loop contribution $Q_{1,2} \times Q_{3-6}$ leads to a significant relative shift of $\Delta\Gamma_s$ as compared to
the 1-loop result. This can be seen from building the ratio between full $\Delta\Gamma_s$
and the $Q_{1,2} \times Q_{3-6}$ piece only. For the width difference incorporating contributions
listed in Table 1 of \cite{Gerlach:2021xtb} and the 1-loop result for $Q_{1,2} \times Q_{3-6}$ we find
\begin{equation}
	\frac{\Delta\Gamma_s^{p,12\times36,\alpha_s^0}}{\Delta\Gamma_s}
	= 7.6 \% \quad ({\rm pole}),\quad 
	\frac{\Delta\Gamma_s^{p,12\times36,\alpha_s^0}}{\Delta\Gamma_s}
	= 6.1 \% \quad (\overline{\rm MS}),
\end{equation}
while the inclusion of the 2-loop $Q_{1,2} \times Q_{3-6}$ piece yields
\begin{equation}
	\frac{\Delta\Gamma_s^{p,12\times36,\alpha_s}}{\Delta\Gamma_s}
	= 0.3 \% \quad ({\rm pole}), \quad
	\frac{\Delta\Gamma_s^{p,12\times36,\alpha_s}}{\Delta\Gamma_s}
	= 1.4 \% \quad (\overline{\rm MS}).
\end{equation}
Here we would like to refer to \cite{Gerlach:2021xtb} for explicit values of all numerical
parameters entering this comparison. 
The notions ``$\overline{\text{MS}}$'' and ``pole'' concern the treatment of the $m_b^2$ prefactor
in Eq.\,\eqref{eq:DB2}. The former means that it is evaluated in the $\overline{\text{MS}}$
scheme, while the latter implies the usage of the on-shell scheme. Notice that even in the pole
scheme all quantities except for the $m_b^2$ prefactor are handled in the $\overline{\text{MS}}$ scheme.

\section{Summary}

In our quest to improve theory prediction for the width difference 
$\Delta \Gamma_s$ in $B^0_s - \bar{B}^0_s$ oscillations we addressed the problem of uncalculated 
QCD corrections at 2- and 3-loop accuracy. The evaluation of these corrections is a crucial step required to achieve a significant 
reduction of the existing perturbative uncertainties. In our matching calculation between $|\Delta B|=1$ and $|\Delta B|=2$ effective theories we were able to obtain fully analytic results for all of the required contributions by
expanding the Feynman diagrams in the ratio $z \equiv m_c^2/m_b^2$. Our final results are valid up to $\mathcal{O}(z)$ at 2 loops and $\mathcal{O}(z^0)$ at 3 loops, while the inclusion of higher orders in $z$ is planned for future iterations
of this work. The first part of our results was made public in \cite{Gerlach:2021xtb}, while the formulas addressing the
remaining 2- and 3-loop contributions will appear in subsequent publications \cite{Gerlach:2021prep2L,Gerlach:2021prep3L}.
In \cite{Gerlach:2021prep3L} we also intend to provide a new theory update on the value $\Delta \Gamma_s$ featuring
NNLO accuracy and reduced theoretical errors.

\section*{Acknowledgments}

We thank Artyom Hovhannisyan for providing to us intermediate results of \cite{Asatrian:2017qaz}. Erik Panzer, Oliver Schnetz, Gudrun Heinrich and Alexander Smirnov are acknowledged for helping with the usage of 
\textsc{HyperInt}, \textsc{HyperLogProcedures}, \textsc{pySecDec} and \textsc{FIESTA} respectively. This
research was supported by the Deutsche Forschungsgemeinschaft (DFG, German Research
 Foundation) under grant 396021762 — TRR 257 “Particle Physics Phenomenology after
the Higgs Discovery”. This report has been assigned preprint numbers TTP21-042 and P3H-21-080.

\bibliographystyle{JHEP}
\bibliography{eps-bmixing.bib}

\providecommand{\href}[2]{#2}\begingroup\raggedright\begin{thebibliography}{10}

\bibitem{Lenz:2021bkv}
A.~Lenz, \emph{{Theory Motivation: What measurements are needed?}},  in
  \emph{{15th International Conference on Heavy Quarks and Leptons}}, 10, 2021
  [\href{https://arxiv.org/abs/2110.01662}{{\ttfamily 2110.01662}}].

\bibitem{LHCb:2019nin}
{\scshape LHCb} collaboration, \emph{{Updated measurement of time-dependent
  {\textbackslash{}it CP}-violating observables in $B^{0}_{s}\to J/\psi K^+
  K^-$ decays}},
  \href{https://doi.org/10.1140/epjc/s10052-019-7159-8}{\emph{Eur. Phys. J. C}
  {\bfseries 79} (2019) 706}
  [\href{https://arxiv.org/abs/1906.08356}{{\ttfamily 1906.08356}}].

\bibitem{CMS:2020efq}
{\scshape CMS} collaboration, \emph{{Measurement of the $CP$-violating phase
  $\phi_\mathrm{s}$ in the B$^0_\mathrm{s}\to$ J$/\psi\, \phi$(1020) $\to
  \mu^+\mu^-$K$^+$K$^-$ channel in proton-proton collisions at $\sqrt{s} =$ 13
  TeV}}, \href{https://doi.org/10.1016/j.physletb.2021.136188}{\emph{Phys.
  Lett. B} {\bfseries 816} (2021) 136188}
  [\href{https://arxiv.org/abs/2007.02434}{{\ttfamily 2007.02434}}].

\bibitem{ATLAS:2020lbz}
{\scshape ATLAS} collaboration, \emph{{Measurement of the $CP$-violating phase
  $\phi_s$ in $B^0_s \to J/\psi\phi$ decays in ATLAS at 13 TeV}},
  \href{https://doi.org/10.1140/epjc/s10052-021-09011-0}{\emph{Eur. Phys. J. C}
  {\bfseries 81} (2021) 342}
  [\href{https://arxiv.org/abs/2001.07115}{{\ttfamily 2001.07115}}].

\bibitem{HFLAV:2019otj}
{\scshape HFLAV} collaboration, \emph{{Averages of b-hadron, c-hadron, and
  $\tau $-lepton properties as of 2018}},
  \href{https://doi.org/10.1140/epjc/s10052-020-8156-7}{\emph{Eur. Phys. J. C}
  {\bfseries 81} (2021) 226}
  [\href{https://arxiv.org/abs/1909.12524}{{\ttfamily 1909.12524}}].

\bibitem{Beneke:1998sy}
M.~Beneke, G.~Buchalla, C.~Greub, A.~Lenz and U.~Nierste,
  \emph{{Next-to-leading order QCD corrections to the lifetime difference of
  B(s) mesons}},
  \href{https://doi.org/10.1016/S0370-2693(99)00684-X}{\emph{Phys. Lett. B}
  {\bfseries 459} (1999) 631}
  [\href{https://arxiv.org/abs/hep-ph/9808385}{{\ttfamily hep-ph/9808385}}].

\bibitem{Ciuchini:2001vx}
M.~Ciuchini, E.~Franco, V.~Lubicz and F.~Mescia, \emph{{Next-to-leading order
  QCD corrections to spectator effects in lifetimes of beauty hadrons}},
  \href{https://doi.org/10.1016/S0550-3213(02)00006-8}{\emph{Nucl. Phys. B}
  {\bfseries 625} (2002) 211}
  [\href{https://arxiv.org/abs/hep-ph/0110375}{{\ttfamily hep-ph/0110375}}].

\bibitem{Ciuchini:2003ww}
M.~Ciuchini, E.~Franco, V.~Lubicz, F.~Mescia and C.~Tarantino, \emph{{Lifetime
  differences and CP violation parameters of neutral B mesons at the
  next-to-leading order in QCD}},
  \href{https://doi.org/10.1088/1126-6708/2003/08/031}{\emph{JHEP} {\bfseries
  08} (2003) 031} [\href{https://arxiv.org/abs/hep-ph/0308029}{{\ttfamily
  hep-ph/0308029}}].

\bibitem{Lenz:2006hd}
A.~Lenz and U.~Nierste, \emph{{Theoretical update of $B_s - \bar{B}_s$
  mixing}}, \href{https://doi.org/10.1088/1126-6708/2007/06/072}{\emph{JHEP}
  {\bfseries 06} (2007) 072}
  [\href{https://arxiv.org/abs/hep-ph/0612167}{{\ttfamily hep-ph/0612167}}].

\bibitem{Asatrian:2017qaz}
H.M.~Asatrian, A.~Hovhannisyan, U.~Nierste and A.~Yeghiazaryan, \emph{{Towards
  next-to-next-to-leading-log accuracy for the width difference in the
  $B_s-\bar{B}_s$ system: fermionic contributions to order $(m_c/m_b)^0$ and
  $(m_c/m_b)^1$}}, \href{https://doi.org/10.1007/JHEP10(2017)191}{\emph{JHEP}
  {\bfseries 10} (2017) 191}
  [\href{https://arxiv.org/abs/1709.02160}{{\ttfamily 1709.02160}}].

\bibitem{Asatrian:2020zxa}
H.M.~Asatrian, H.H.~Asatryan, A.~Hovhannisyan, U.~Nierste, S.~Tumasyan and
  A.~Yeghiazaryan, \emph{{Penguin contribution to the width difference and $CP$
  asymmetry in $B_q$-$\bar B_q$ mixing at order $\alpha_s^2 N_f$}},
  \href{https://doi.org/10.1103/PhysRevD.102.033007}{\emph{Phys. Rev. D}
  {\bfseries 102} (2020) 033007}
  [\href{https://arxiv.org/abs/2006.13227}{{\ttfamily 2006.13227}}].

\bibitem{Chetyrkin:1997gb}
K.G.~Chetyrkin, M.~Misiak and M.~Munz, \emph{{$|\Delta F| = 1$ nonleptonic
  effective Hamiltonian in a simpler scheme}},
  \href{https://doi.org/10.1016/S0550-3213(98)00131-X}{\emph{Nucl. Phys. B}
  {\bfseries 520} (1998) 279}
  [\href{https://arxiv.org/abs/hep-ph/9711280}{{\ttfamily hep-ph/9711280}}].

\bibitem{Dugan:1990df}
M.J.~Dugan and B.~Grinstein, \emph{{On the vanishing of evanescent operators}},
  \href{https://doi.org/10.1016/0370-2693(91)90680-O}{\emph{Phys. Lett. B}
  {\bfseries 256} (1991) 239}.

\bibitem{Herrlich:1994kh}
S.~Herrlich and U.~Nierste, \emph{{Evanescent operators, scheme dependences and
  double insertions}},
  \href{https://doi.org/10.1016/0550-3213(95)00474-7}{\emph{Nucl. Phys. B}
  {\bfseries 455} (1995) 39}
  [\href{https://arxiv.org/abs/hep-ph/9412375}{{\ttfamily hep-ph/9412375}}].

\bibitem{Gerlach:2021xtb}
M.~Gerlach, U.~Nierste, V.~Shtabovenko and M.~Steinhauser, \emph{{Two-loop QCD
  penguin contribution to the width difference in B$_{s}$\ensuremath{-}$
  {\overline{B}}_s $ mixing}},
  \href{https://doi.org/10.1007/JHEP07(2021)043}{\emph{JHEP} {\bfseries 07}
  (2021) 043} [\href{https://arxiv.org/abs/2106.05979}{{\ttfamily
  2106.05979}}].

\bibitem{Gerlach:2021prep3L}
M.~Gerlach, U.~Nierste, V.~Shtabovenko and M.~Steinhauser, \emph{{NNLO
  corrections to the width difference in B$_{s}$\ensuremath{-}$
  {\overline{B}}_s $ mixing}}, {\emph{in preparation} }.

\bibitem{Khoze:1983yp}
V.A.~Khoze and M.A.~Shifman, \emph{{HEAVY QUARKS}},
  \href{https://doi.org/10.1070/PU1983v026n05ABEH004398}{\emph{Sov. Phys. Usp.}
  {\bfseries 26} (1983) 387}.

\bibitem{Shifman:1984wx}
M.A.~Shifman and M.B.~Voloshin, \emph{{Preasymptotic Effects in Inclusive Weak
  Decays of Charmed Particles}}, {\emph{Sov. J. Nucl. Phys.} {\bfseries 41}
  (1985) 120}.

\bibitem{Khoze:1986fa}
V.A.~Khoze, M.A.~Shifman, N.G.~Uraltsev and M.B.~Voloshin, \emph{{On Inclusive
  Hadronic Widths of Beautiful Particles}}, {\emph{Sov. J. Nucl. Phys.}
  {\bfseries 46} (1987) 112}.

\bibitem{Chay:1990da}
J.~Chay, H.~Georgi and B.~Grinstein, \emph{{Lepton energy distributions in
  heavy meson decays from QCD}},
  \href{https://doi.org/10.1016/0370-2693(90)90916-T}{\emph{Phys. Lett. B}
  {\bfseries 247} (1990) 399}.

\bibitem{Bigi:1991ir}
I.I.Y.~Bigi and N.G.~Uraltsev, \emph{{Gluonic enhancements in non-spectator
  beauty decays: An Inclusive mirage though an exclusive possibility}},
  \href{https://doi.org/10.1016/0370-2693(92)90066-D}{\emph{Phys. Lett. B}
  {\bfseries 280} (1992) 271}.

\bibitem{Bigi:1992su}
I.I.Y.~Bigi, N.G.~Uraltsev and A.I.~Vainshtein, \emph{{Nonperturbative
  corrections to inclusive beauty and charm decays: QCD versus phenomenological
  models}}, \href{https://doi.org/10.1016/0370-2693(92)90908-M}{\emph{Phys.
  Lett. B} {\bfseries 293} (1992) 430}
  [\href{https://arxiv.org/abs/hep-ph/9207214}{{\ttfamily hep-ph/9207214}}].

\bibitem{Bigi:1993fe}
I.I.Y.~Bigi, M.A.~Shifman, N.G.~Uraltsev and A.I.~Vainshtein, \emph{{QCD
  predictions for lepton spectra in inclusive heavy flavor decays}},
  \href{https://doi.org/10.1103/PhysRevLett.71.496}{\emph{Phys. Rev. Lett.}
  {\bfseries 71} (1993) 496}
  [\href{https://arxiv.org/abs/hep-ph/9304225}{{\ttfamily hep-ph/9304225}}].

\bibitem{Blok:1993va}
B.~Blok, L.~Koyrakh, M.A.~Shifman and A.I.~Vainshtein, \emph{{Differential
  distributions in semileptonic decays of the heavy flavors in QCD}},
  \href{https://doi.org/10.1103/PhysRevD.50.3572}{\emph{Phys. Rev. D}
  {\bfseries 49} (1994) 3356}
  [\href{https://arxiv.org/abs/hep-ph/9307247}{{\ttfamily hep-ph/9307247}}].

\bibitem{Manohar:1993qn}
A.V.~Manohar and M.B.~Wise, \emph{{Inclusive semileptonic B and polarized
  Lambda(b) decays from QCD}},
  \href{https://doi.org/10.1103/PhysRevD.49.1310}{\emph{Phys. Rev. D}
  {\bfseries 49} (1994) 1310}
  [\href{https://arxiv.org/abs/hep-ph/9308246}{{\ttfamily hep-ph/9308246}}].

\bibitem{Nogueira:1991ex}
P.~Nogueira, \emph{{Automatic Feynman graph generation}},
  \href{https://doi.org/10.1006/jcph.1993.1074}{\emph{J. Comput. Phys.}
  {\bfseries 105} (1993) 279}.

\bibitem{Harlander:1998cmq}
R.~Harlander, T.~Seidensticker and M.~Steinhauser, \emph{{Complete corrections
  of Order alpha alpha-s to the decay of the Z boson into bottom quarks}},
  \href{https://doi.org/10.1016/S0370-2693(98)00220-2}{\emph{Phys. Lett. B}
  {\bfseries 426} (1998) 125}
  [\href{https://arxiv.org/abs/hep-ph/9712228}{{\ttfamily hep-ph/9712228}}].

\bibitem{Seidensticker:1999bb}
T.~Seidensticker, \emph{{Automatic application of successive asymptotic
  expansions of Feynman diagrams}},  in \emph{{6th International Workshop on
  New Computing Techniques in Physics Research: Software Engineering,
  Artificial Intelligence Neural Nets, Genetic Algorithms, Symbolic Algebra,
  Automatic Calculation}}, 5, 1999
  [\href{https://arxiv.org/abs/hep-ph/9905298}{{\ttfamily hep-ph/9905298}}].

\bibitem{Gerlach:tapir}
M.~Gerlach and F.~Herren, \emph{in preparation}, .

\bibitem{Kuipers:2012rf}
J.~Kuipers, T.~Ueda, J.A.M.~Vermaseren and J.~Vollinga, \emph{{FORM version
  4.0}}, \href{https://doi.org/10.1016/j.cpc.2012.12.028}{\emph{Comput. Phys.
  Commun.} {\bfseries 184} (2013) 1453}
  [\href{https://arxiv.org/abs/1203.6543}{{\ttfamily 1203.6543}}].

\bibitem{Alloul:2013bka}
A.~Alloul, N.D.~Christensen, C.~Degrande, C.~Duhr and B.~Fuks, \emph{{FeynRules
  2.0 - A complete toolbox for tree-level phenomenology}},
  \href{https://doi.org/10.1016/j.cpc.2014.04.012}{\emph{Comput. Phys. Commun.}
  {\bfseries 185} (2014) 2250}
  [\href{https://arxiv.org/abs/1310.1921}{{\ttfamily 1310.1921}}].

\bibitem{Hahn:2000kx}
T.~Hahn, \emph{{Generating Feynman diagrams and amplitudes with FeynArts 3}},
  \href{https://doi.org/10.1016/S0010-4655(01)00290-9}{\emph{Comput. Phys.
  Commun.} {\bfseries 140} (2001) 418}
  [\href{https://arxiv.org/abs/hep-ph/0012260}{{\ttfamily hep-ph/0012260}}].

\bibitem{Mertig:1990an}
R.~Mertig, M.~Bohm and A.~Denner, \emph{{FEYN CALC: Computer algebraic
  calculation of Feynman amplitudes}},
  \href{https://doi.org/10.1016/0010-4655(91)90130-D}{\emph{Comput. Phys.
  Commun.} {\bfseries 64} (1991) 345}.

\bibitem{Shtabovenko:2016sxi}
V.~Shtabovenko, R.~Mertig and F.~Orellana, \emph{{New Developments in FeynCalc
  9.0}}, \href{https://doi.org/10.1016/j.cpc.2016.06.008}{\emph{Comput. Phys.
  Commun.} {\bfseries 207} (2016) 432}
  [\href{https://arxiv.org/abs/1601.01167}{{\ttfamily 1601.01167}}].

\bibitem{Shtabovenko:2020gxv}
V.~Shtabovenko, R.~Mertig and F.~Orellana, \emph{{FeynCalc 9.3: New features
  and improvements}},
  \href{https://doi.org/10.1016/j.cpc.2020.107478}{\emph{Comput. Phys. Commun.}
  {\bfseries 256} (2020) 107478}
  [\href{https://arxiv.org/abs/2001.04407}{{\ttfamily 2001.04407}}].

\bibitem{Lewis:Fermat}
R.~Lewis, \emph{{FERMAT}}, {\emph{\url{https://home.bway.net/lewis}} }.

\bibitem{Pak:2011xt}
A.~Pak, \emph{{The Toolbox of modern multi-loop calculations: novel analytic
  and semi-analytic techniques}},
  \href{https://doi.org/10.1088/1742-6596/368/1/012049}{\emph{J. Phys. Conf.
  Ser.} {\bfseries 368} (2012) 012049}
  [\href{https://arxiv.org/abs/1111.0868}{{\ttfamily 1111.0868}}].

\bibitem{Smirnov:2019qkx}
A.V.~Smirnov and F.S.~Chuharev, \emph{{FIRE6: Feynman Integral REduction with
  Modular Arithmetic}},
  \href{https://doi.org/10.1016/j.cpc.2019.106877}{\emph{Comput. Phys. Commun.}
  {\bfseries 247Â } (2020) 106877}
  [\href{https://arxiv.org/abs/1901.07808}{{\ttfamily 1901.07808}}].

\bibitem{Lee:2013mka}
R.N.~Lee, \emph{{LiteRed 1.4: a powerful tool for reduction of multiloop
  integrals}}, \href{https://doi.org/10.1088/1742-6596/523/1/012059}{\emph{J.
  Phys. Conf. Ser.} {\bfseries 523} (2014) 012059}
  [\href{https://arxiv.org/abs/1310.1145}{{\ttfamily 1310.1145}}].

\bibitem{Chetyrkin:1981qh}
K.G.~Chetyrkin and F.V.~Tkachov, \emph{{Integration by Parts: The Algorithm to
  Calculate beta Functions in 4 Loops}},
  \href{https://doi.org/10.1016/0550-3213(81)90199-1}{\emph{Nucl. Phys. B}
  {\bfseries 192} (1981) 159}.

\bibitem{Tkachov:1981wb}
F.V.~Tkachov, \emph{{A Theorem on Analytical Calculability of Four Loop
  Renormalization Group Functions}},
  \href{https://doi.org/10.1016/0370-2693(81)90288-4}{\emph{Phys. Lett. B}
  {\bfseries 100} (1981) 65}.

\bibitem{Smirnov:2012gma}
V.A.~Smirnov, \emph{{Analytic tools for Feynman integrals}}, vol.~250 (2012),
  \href{https://doi.org/10.1007/978-3-642-34886-0}{10.1007/978-3-642-34886-0}.

\bibitem{Fleischer:1999hp}
J.~Fleischer, M.Y.~Kalmykov and A.V.~Kotikov, \emph{{Two loop selfenergy master
  integrals on-shell}},
  \href{https://doi.org/10.1016/S0370-2693(99)00892-8}{\emph{Phys. Lett. B}
  {\bfseries 462} (1999) 169}
  [\href{https://arxiv.org/abs/hep-ph/9905249}{{\ttfamily hep-ph/9905249}}].

\bibitem{Panzer:2014caa}
E.~Panzer, \emph{{Algorithms for the symbolic integration of hyperlogarithms
  with applications to Feynman integrals}},
  \href{https://doi.org/10.1016/j.cpc.2014.10.019}{\emph{Comput. Phys. Commun.}
  {\bfseries 188} (2015) 148}
  [\href{https://arxiv.org/abs/1403.3385}{{\ttfamily 1403.3385}}].

\bibitem{Schnetz:HLP}
O.~Schnetz, \emph{{HyperLogProcedures}},
  {\emph{\url{https://www.math.fau.de/person/oliver-schnetz}} }.

\bibitem{Duhr:2019tlz}
C.~Duhr and F.~Dulat, \emph{{PolyLogTools \textemdash{} polylogs for the
  masses}}, \href{https://doi.org/10.1007/JHEP08(2019)135}{\emph{JHEP}
  {\bfseries 08} (2019) 135}
  [\href{https://arxiv.org/abs/1904.07279}{{\ttfamily 1904.07279}}].

\bibitem{Borowka:2017idc}
S.~Borowka, G.~Heinrich, S.~Jahn, S.P.~Jones, M.~Kerner, J.~Schlenk et~al.,
  \emph{{pySecDec: a toolbox for the numerical evaluation of multi-scale
  integrals}}, \href{https://doi.org/10.1016/j.cpc.2017.09.015}{\emph{Comput.
  Phys. Commun.} {\bfseries 222} (2018) 313}
  [\href{https://arxiv.org/abs/1703.09692}{{\ttfamily 1703.09692}}].

\bibitem{Borowka:2018goh}
S.~Borowka, G.~Heinrich, S.~Jahn, S.P.~Jones, M.~Kerner and J.~Schlenk,
  \emph{{A GPU compatible quasi-Monte Carlo integrator interfaced to
  pySecDec}}, \href{https://doi.org/10.1016/j.cpc.2019.02.015}{\emph{Comput.
  Phys. Commun.} {\bfseries 240} (2019) 120}
  [\href{https://arxiv.org/abs/1811.11720}{{\ttfamily 1811.11720}}].

\bibitem{Heinrich:2021dbf}
G.~Heinrich, S.~Jahn, S.P.~Jones, M.~Kerner, F.~Langer, V.~Magerya et~al.,
  \emph{{Expansion by regions with pySecDec}},
  \href{https://arxiv.org/abs/2108.10807}{{\ttfamily 2108.10807}}.

\bibitem{Smirnov:2015mct}
A.V.~Smirnov, \emph{{FIESTA4: Optimized Feynman integral calculations with GPU
  support}}, \href{https://doi.org/10.1016/j.cpc.2016.03.013}{\emph{Comput.
  Phys. Commun.} {\bfseries 204} (2016) 189}
  [\href{https://arxiv.org/abs/1511.03614}{{\ttfamily 1511.03614}}].

\bibitem{Gambino:2003zm}
P.~Gambino, M.~Gorbahn and U.~Haisch, \emph{{Anomalous dimension matrix for
  radiative and rare semileptonic B decays up to three loops}},
  \href{https://doi.org/10.1016/j.nuclphysb.2003.09.024}{\emph{Nucl. Phys. B}
  {\bfseries 673} (2003) 238}
  [\href{https://arxiv.org/abs/hep-ph/0306079}{{\ttfamily hep-ph/0306079}}].

\bibitem{Beneke:1996gn}
M.~Beneke, G.~Buchalla and I.~Dunietz, \emph{{Width Difference in the
  $B_s-\bar{B_s}$ System}},
  \href{https://doi.org/10.1103/PhysRevD.54.4419}{\emph{Phys. Rev. D}
  {\bfseries 54} (1996) 4419}
  [\href{https://arxiv.org/abs/hep-ph/9605259}{{\ttfamily hep-ph/9605259}}].

\bibitem{Gerlach:2021prep2L}
M.~Gerlach, U.~Nierste, V.~Shtabovenko and M.~Steinhauser, \emph{{Complete NLO
  and partial NNLO corrections to the width difference in
  B$_{s}$\ensuremath{-}$ {\overline{B}}_s $ mixing}}, {\emph{in preparation} }.

\end{thebibliography}\endgroup


\end{document}